\documentclass[aps,prl,twocolumn,superscriptaddress,showpacs]{revtex4}
\usepackage{amsfonts,amssymb,amsmath,latexsym,epsfig} 

\begin{document}

\title{Poincar\'e recurrences from the perspective of transient chaos \\ 
{\small Phys. Rev. Lett. 100, 174101 (2008)}}
\date{\today}
\author{Eduardo G. Altmann }
\altaffiliation{Current address: Northwestern Institute on Complex Systems, Northwestern University, Evanston, IL, USA \\ E-mail: ega@northwestern.edu}

\affiliation{Max Planck Institute for the Physics of Complex Systems, 01187 Dresden, Germany}

\author{Tam\'as T\'el}

\affiliation{Institute for Theoretical Physics, E\"otv\"os University, P.O. Box 32, H-1518 Budapest, Hungary}

\begin{abstract}
 We obtain a description of the Poincar\'e recurrences of chaotic systems in
 terms of the ergodic theory of transient chaos. It is based on the equivalence between
 the recurrence time 
 distribution and an escape time distribution obtained by leaking the
 system and taking a special initial ensemble. This ensemble is atypical in terms
 of the  natural measure of the leaked system, the  conditionally invariant
 measure. Accordingly, for general initial ensembles, the average recurrence and
 escape times are different. However, we show that the decay rate of 
 these distributions is always the same. Our results remain valid for
 Hamiltonian systems with mixed phase space and validate a split of the
 chaotic saddle in hyperbolic and non-hyperbolic  components. 

\end{abstract}

\maketitle

The idea of recurrence to a certain region of the phase space
was introduced by Poincar\'e in his famous studies of the three body problem
as a stability criterion~\cite{poincare}. The wider implications of his results were soon
recognized and Poincar\'e recurrences have played an important role ever since
the first debates on the foundation of nonequilibrium
processes~\cite{kac,Dorfman,Zasl1}. More recently, recurrences have become a
standard tool to investigate low-dimensional {\em closed} chaotic systems 
\cite{CS,ZT,Zasl2,Hadyn,maths,maths2,altmann,BKG,ketzmerick}.

{\em Open} systems, on the other hand, are usually investigated 
as scattering problems: the long-lasting chaotic transients are
related to an invariant fractal saddle~\cite{LFO,ott,tel} and only 
short time escapes depend on the initial ensemble. 
A sharp distinction between open and closed systems exists since the time of
Poincar\'e~\cite{poincare}.    

Closed systems can be converted into open ones by defining
a finite region of the phase space as {\em a leak}. 
Leaking dynamical systems mimics the effect of experimental
observations~\cite{DS,BD,optics} and has also been applied as a tool to investigate
the dynamics of closed systems~\cite{pierrehumbert,paar,schneider}.

In this Letter, we 
find the complete correspondence between the recurrence
and the leak problems. We then
apply the transient chaos theory of
open systems to describe Poincar\'e recurrences of closed systems. More
precisely, {\em considering the recurrence region 
  to be the leak},
we show that for chaotic (both dissipative and Hamiltonian) systems (i) 
the exponential relaxation rate of the recurrence problem and the escape rate
from the leaked system are always identical, 
(ii) the relaxation rate can be expressed by the conditionally invariant measure \cite{PY} of the leaked region, (iii) 
under a properly chosen initial ensemble
of the escape problem, the entire recurrence and escape time distributions coincide,
and (iv) the points above remain valid even in Hamiltonian systems with
mixed phase space, where for recurrence/leak regions far from 
Kolmogorov-Arnold-Moser (KAM) islands~\cite{ott,tel} the escape rate and an associated hyperbolic saddle are shown to
be well defined for times shorter than a crossover time.

To be specific, consider a chaotic map $M$ defined on a bounded phase space
$\Gamma$ 
with an invariant chaotic set: a chaotic attractor or a chaotic sea.
The natural ergodic measure on this invariant set is denoted by $\mu$,
its density by $\rho_\mu$. The recurrence region is a subset $I\subset\Gamma$. We
{\em do not} restrict ourselves to small~$\mu(I)$~\cite{altmann,BKG,maths2}.
The Poincar\'e  recurrence theorem \cite{Zasl1,Dorfman,kac} 
ensures that for almost any 
initial condition in $I$, 
there is a first recurrence
at some discrete time $n_1$, a second one at time $n_1+n_2$, etc. In
the long time limit, the set of recurrence times~$n_i$ defines the
recurrence time distribution $p_r(n)$.   
For chaotic systems \cite{maths,Zasl1,altmann}, this distribution 
is exponential for $n$ larger than some small $n^*_r$
\begin{equation}\label{eq.expr}
p_r(n) \approx g_r e^{-\gamma_r n}.
\end{equation}
Prefactor $g_r$ and the relaxation rate $\gamma_r$ depend on the choice
of~$I$.   
While the average recurrence time $\langle n \rangle_r$ is given by Kac's lemma
\cite{kac}:  
\begin{equation}\label{eq.kac}
\langle n \rangle_r \equiv \sum_{n=1}^\infty n p_r(n) = \frac{1}{\mu(I)},
\end{equation}
it has been unknown whether the relaxation rate~$\gamma_r$ can be expressed by
means of any measure of~$I$. 
The equality $\gamma_r=1/\langle n_r \rangle = \mu(I)$ is proved only in the
unrealistic limit $\mu(I)\rightarrow 0$ (see \cite{maths}, and references
therein).  

Open (transiently) chaotic maps are characterized by the escape time
distribution $p_e(n)$, which gives the fraction of trajectories of a
certain initial ensemble that leaves the system at time $n$. This distribution is
also exponential \cite{ott,tel}. We are interested in the class of
open systems obtained by leaking a closed system at a region~$I\subset\Gamma$,
described by map $M^\dagger(\vec{x})=M(\vec{x})$ for $\vec{x} \in \Gamma \setminus I$
and~$M^\dagger(\vec{x})=$`exit' for $\vec{x} \in I$. Notice that the escape
happens one step {\em after} entering~$I$ and initial conditions can be in $I$. In particular, for such a leaked system we have
\begin{equation}\label{eq.expe}
p_e(n) \approx g_e e^{-\gamma_e n},
\end{equation}
for $n \ge n^*_e$, where $\gamma_e$ is the escape rate. The
average escape time $\langle n \rangle_e$, also called the lifetime of chaos,
is usually estimated by the reciprocal of $\gamma_e$ \cite{ott,tel}. 
While the full $p_e(n)$, and the
average $\langle n \rangle_e$, depend also on the choice of the {\em initial density} $\rho_0$ used in generating $p_e(n)$, the escape rate does not
(provided it exists). All these quantities depend on $I$. The
theory of transient chaos \cite{ott,tel} explains these results based on the
existence of an invariant chaotic saddle {\em in the complement set  of the
  leak $\Gamma \setminus  I$}~\cite{notenew}. The saddle is the union of all trajectories
that 
{\em never} enter the leak, neither forward nor backward in time. It is a
fractal set of measure zero for
strongly chaotic systems.
Trajectories spending a long time outside~$I$ must come close to the saddle along
its stable manifold, stay in the vicinity of the saddle, and leave it along
its unstable manifold. Therefore, irrespective of $\rho_0$, the
long-term emptying process is governed by the invariant saddle, which is
uniquely defined by~$I$. 

\begin{figure}[!ht]
\includegraphics[width=1\columnwidth]{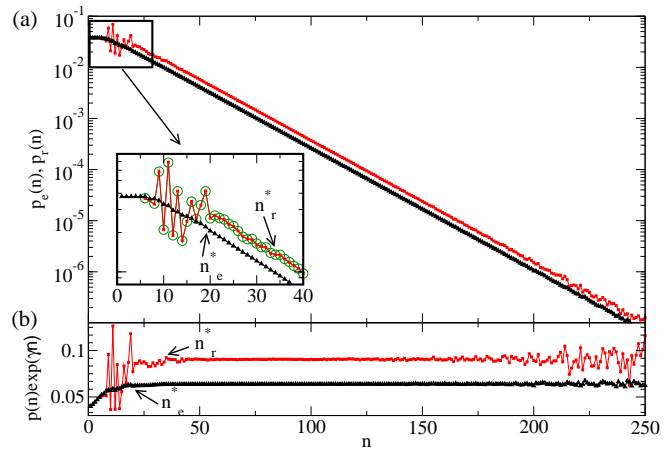}
\caption{ (Color online)
 (a) Recurrence $p_r(n)$ ({\tiny
 $\blacksquare$}/above) and escape $p_e(n)$ ($ \blacktriangle$/below)  time distributions in the
H\'enon map shown in Fig.~\ref{fig2}. For $p_r(n)$   a single initial condition
 $(x_0,y_0)=(x_c,y_c)$ was iterated over $10^{11}$ steps. For  $p_e(n)$,
 $\rho_0=\rho_\mu$ was built by~$7\cdot10^9$ trajectories. Inset: short time
 behavior, where $p_e(n)$ with $\rho_0=\rho_r$, given
by~(\ref{eq.rhor}), is marked by {\Large $\circ$}. 
(b) $p_{e,r} \exp(\gamma n)$,
with~$\gamma=0.055$ [$\mu_c(I)=0.0535 < \mu(I)=0.0377$]. Numerical limitations
 appear for large~$n$.   
}  
\label{fig1}
\end{figure}

\begin{figure}[!ht]
\includegraphics[width=1\columnwidth]{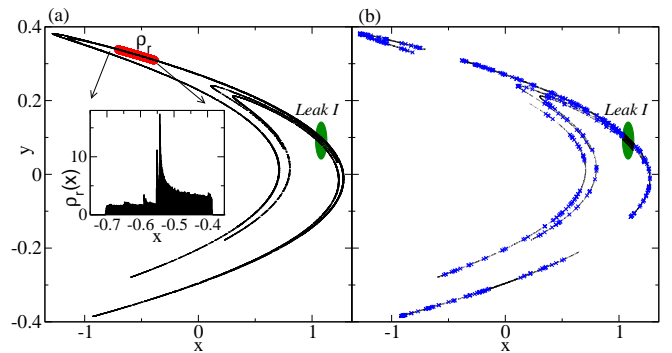}
\caption{(Color online) Phase space of the H\'enon map
  $(x_{n+1},y_{n+1})=(1-1.4 x_n^2+y_n,0.3 x_{n})$ with $I$ as a circle of
  radius $0.05$ centered at~$(x_c,y_c)=(1.0804,0.0916)$.
  (a) H\'enon attractor: support of $\rho_\mu$ (thin line) and of $\rho_r$
  ({\Large $\bullet$}) at~$M(I)$, see~(\ref{eq.rhor}). Inset: $\rho_r(x,y)$ projected
  onto $x$. (b) Chaotic saddle $(\times)$ 
  and the support of 
$\rho_c$  (thin line).  
}
\label{fig2}
\end{figure}

We now apply this picture to Poincar\'e recurrences. 
Long recurrences should correspond to trajectories that fall near to the chaotic saddle's stable manifold right after 
exiting the recurrence region $I$. 
Therefore, we expect the relaxation rate of the recurrence
distributions~$\gamma_r$ and the decay rate of the escape
distribution~$\gamma_e$ to coincide 
\begin{equation}\label{eq.gammare}
\gamma_r=\gamma_e\equiv\gamma,
\end{equation}
for a given recurrence/leak region~$I$.
This equality is illustrated for the H\'enon map in Fig.~\ref{fig1},
where it is apparent that the slopes of $\ln{p_r(n)}$ vs. $n$
and  $\ln{p_e(n)}$ vs. $n$ coincide. 
Equation (\ref{eq.gammare}), which will be rigorously justified below, implies that to
any Poincar\'e recurrence problem there exists a chaotic saddle (that of the
corresponding  leaked system) that governs the long-term recurrences and
determines $\gamma_r$. This saddle is shown in Fig.~\ref{fig2}b. Its double
fractal character
 makes evident the improvement of 
Eq.~(\ref{eq.gammare}) in comparison to the previous results being limited to $\mu(I)\rightarrow 0$~\cite{Zasl2,maths}.

Despite having the same exponential decay~(\ref{eq.gammare}), 
$p_r(n)$ and $p_e(n)$ are different. This holds 
for practically any initial density $\rho_0$. We show now that there is,
nevertheless, a special $\rho_0=\rho_r$~\cite{maths,maths2}
of the leaked system for which $p_e(n)=p_r(n)$.
Consider inserting trajectories into the system through the
leak. More precisely, consider the density~$\rho_r(\vec{x})$ obtained as the
first iterate  through  map~$M$ of the points $\vec{x} \in I$ distributed
according to the natural density of the closed system $\rho_\mu$: 
\begin{equation}
\rho_r(\vec{x}) = \rho_\mu(M^{-1}(\vec{x})\cap I)/\mu(M(I)) \;\; \text{ for } \;\; \vec{x}\in M(I),
\label{eq.rhor}
\end{equation}
where $M^{-1}(\vec{x}) \cap I$ denotes the points that come from~$I$ and $\mu(M(I))$ ensures normalization. This distribution is
shown in Fig.~\ref{fig2}a.
In view of the ergodic theorem, the distribution of all positions of
a single infinitely long recurrent trajectory, one iteration after returning
to~$I$, is precisely given by $\rho_r$. Since the escape times
for~$\rho_r$ are exactly the recurrence times, it follows that 
$$p_r(n)=p_e(n) \;\;\; \mbox{with} \;\;\; \rho_0=\rho_r,$$
%
%
for all times~$n$ (see the inset of Fig.~\ref{fig1}a). 
Of course $\gamma_r=\gamma_e$ and, because 
$\gamma_e$ is
independent of 
$\rho_0$, this implies~(\ref{eq.gammare}).

Having shown how recurrences can be viewed as escapes from a leaked system, we
now perform a formal description in terms of the ergodic theory of
transient chaos. A central concept in this theory is the so-called
conditionally invariant measure, c-measure in brief~\cite{PY,DY,T}. 
Smooth initial densities iterated via the action of map $M^\dagger$ do not
converge to any finite measure due to the permanent escape. If, however,  one
applies a compensation in the form of multiplying the density in each step by
$\exp{(\gamma)}>1$ it converges to a finite measure, the c-measure
  $\mu_c$. 
Due to the permanent contraction along the stable direction, the density
$\rho_c$ of the c-measure is nonzero along the unstable manifold of the
saddle~\cite{T}.  Qualitatively speaking, this measure characterizes 
the escaping process from the saddle and is thus completely different from the 
natural measure of the closed system. With the normalization
$\mu_c(\Gamma)=1$, the compensation factor~$\exp(\gamma)$ is the c-measure of
the region not escaping within one time step~\cite{PY}. 
In the leaked system this is
$\mu_c(\Gamma \setminus  I)=\mu_c(\Gamma)-\mu_c(I)=1-\mu_c(I)$ and therefore 
\begin{equation}\label{eq.muc}
\gamma=-\ln{(1-\mu_c(I))} \;\;\;\; [ \approx \mu_c(I) \;\; \mbox{for small}
  \;\; \mu(I) ].
\end{equation}
Note that the c-measure itself depends on  the form, size, and location
of~$I$. In view of (\ref{eq.gammare}), it is remarkable that the  
relaxation rate~$\gamma_r$ of the Poincar\'e recurrences of the closed system is given by
the c-measure of the recurrence region, when viewed as a leaked system.
Furthermore, 
$\rho_0=\rho_c$ is the only initial density for which the
exponential decay starts from the very first iterate ($n^*_e=1$).
As a consequence, 
%
\begin{equation*}\label{eq.nmuc}
\langle n \rangle_{e}=1/\mu_c(I) \;\;\;\; \mbox{with} \;\;\;\; \rho_0=\rho_c.
\end{equation*}
The reason for the strong difference between this relation and 
Kac's lemma (\ref{eq.kac}) (valid for recurrences 
and for $\rho_0=\rho_r$) is that $\rho_r$ in (\ref{eq.rhor}) is atypical
from the point of view of the
c-measure. In fact, $\rho_c$ is concentrated along the
saddle's unstable manifold while all points 
on~$\rho_r$ have come from the leak. Typical initial densities~$\rho_0$
quickly converge to $\rho_c$, explaining why typical $\langle n
\rangle_e$-s, but not~$\langle n \rangle_r$, are well estimated 
by~$1/\gamma\approx1/\mu_c(I)$~\cite{toappear}.

We illustrate now through two examples the new perspectives opened by our unified approach. 
(i) Recently, it has been surprisingly observed that different leaks~$I$ with
equal~$\mu(I)$ lead to different $\gamma_e$'s~\cite{paar,schneider}. Location
dependence of $\gamma_r$ has also been reported~\cite{ZT,altmann,BKG}. In view of
(\ref{eq.gammare}), these works
refer to the same
phenomenon. Furthermore, Eq.~(\ref{eq.muc}) 
explains the observations by stating that it is the c-measure and not the
natural measure that determines~$\gamma$. 
(ii) 
A basic problem in the mathematical approach to recurrences is the convergence of $p_r(n)$ to a Poissonian 
[$ \gamma=\mu(I) $] for $\mu(I) \rightarrow 0$~\cite{maths}. 
A
fundamental property of the c-measure is its 
converges to the natural measure \cite{PY,DY}.  
Equation~(\ref{eq.muc}) implies that both convergences occur precisely
in the same way. 

We turn now to Hamiltonian systems with mixed phase space,
where deviations from the exponential decay appear for long times. The
trajectories that approach the KAM islands (or other non-hyperbolic structures) {\em stick} to them
for a time that is roughly algebraically distributed~\cite{CS,Zasl1,ketzmerick}. 
This phenomenon is usually investigated using recurrences~\cite{CS,Zasl1,ketzmerick}
but appears also in escape problems~\cite{tel,pikovsky}.
The asymptotic power-law like decay of $p_{r,e}(n)$ is the same 
and independent of the choice of initial densities (away from
islands~\cite{pikovsky}), because it is related to the non-hyperbolic structures. For shorter times, an exponential decay of $p_{r,e}(n)$ is 
observed~\cite{ZT}, what has deserved little attention from the recurrence perspective
so far.

\begin{figure}[!ht]
\includegraphics[width=1\columnwidth]{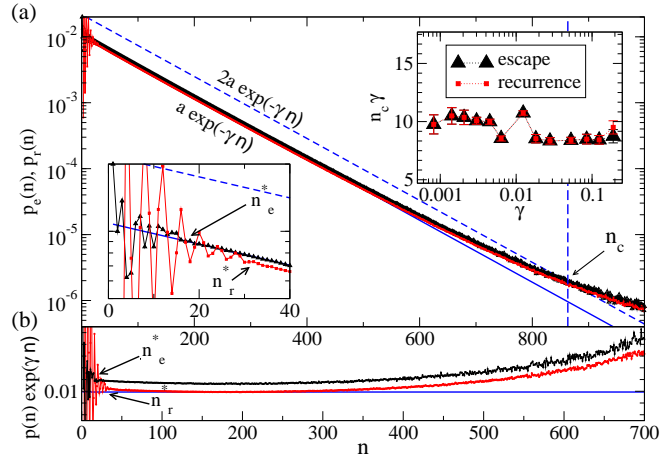}
\caption{ (Color online)
(a) Recurrence $p_r(n)$ ({\tiny $\blacksquare$}/below) and escape $p_e(n)$ ($\blacktriangle$/above) time distributions in the
standard map shown in Fig.~\ref{fig4} with~$\varepsilon=0.1$. For $p_r(n)$ 
a single initial condition  was iterated over $10^{11}$ steps. For $p_e(n)$,
$\rho_0=\rho_\mu$ at $|x|>0.25$ was built by~$2\cdot10^8$ trajectories. Lower
inset: short time behavior. 
Upper inset: scaling of $n_c \gamma$
with~$\gamma$, obtained by changing $\varepsilon$ in $0.03\le \varepsilon\le
0.4$. (b) $p_{e,r}\exp(\gamma n)$, with 
$\gamma=0.0108$ [$\mu_c(I)=0.0107 <\mu(I)=0.0113$]. 
}  
\label{fig3}
\end{figure}

Accordingly, for $I$ outside KAM islands and for $n\ge n^*$ 
\begin{equation}\label{eq.powerlaw}
p_{r,e}(n) \approx \left\{ \begin{array}{ll}
   a e^{-\gamma n} & \text{ for } n^* < n < n_{\alpha} \\
   a e^{-\gamma n} + b (\gamma n)^{-\alpha} & \text{ for } n > n_{\alpha}, \\
\end{array} 
\right.
\end{equation}
where $a e^{-\gamma n_{\alpha}} \gg b \;{(\gamma n_{\alpha})}^{-\alpha}$.
The recurrence and escape problems can be again related
using~$\rho_r$ in~(\ref{eq.rhor}). However,
the analysis based on the chaotic saddle has to be reconsidered 
in the presence of KAM islands. Numerical results for the standard map in Fig.~\ref{fig3} indicate that, for {\em intermediate} times,  
$\gamma_{r,e}$ are well defined and equal, as in~(\ref{eq.gammare}). These
results remain valid~\cite{toappear} for a wide range of~$I$'s and of smooth~$\rho_0$'s, 
away from islands (see~\cite{pikovsky}  
for $\rho_0$'s touching the island).

\begin{figure}[!ht]
\includegraphics[width=1\columnwidth]{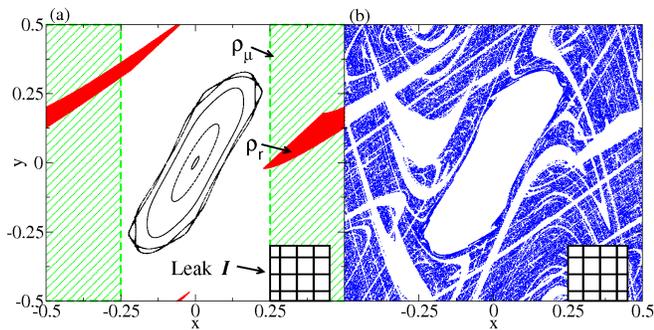}
\caption{ (Color online)
(a) Phase space of the standard map $(x_{n+1},y_{n+1})=(x_n+y_{n+1},y_n-0.52 \sin(2\pi x_n))$, periodic
in $|x,y|<0.5$. $I=[0.25<x<0.25+\varepsilon,-0.5<y<-0.5+\varepsilon]$,
$\varepsilon=0.2$ ($\boxplus$). The
  KAM island (non-hyperbolic component of the saddle) is at the center. The support
  of~$\rho_r$  is at $M(I)$ (dark/red region). (b) The
  hyperbolic component of the saddle generated by the method
  of~\cite{tel}, using~$n=60 < n_c=160$.
}  
\label{fig4}
\end{figure}

To understand these findings we propose here to
effectively split the saddle in a hyperbolic (outside
islands) and a nonhyperbolic (nearly space filling \cite{LFO}) component.
To justify this splitting, it is worth defining a crossover
time $n_c>n_{\alpha}$               
as
$a e^{-\gamma n_c}= b (\gamma n_c)^{-\alpha},$
%
starting from which the nonhyperbolic component dominates. 
In the upper inset of Fig.~\ref{fig3}, we present evidence for
%
\begin{equation}\label{eq.nc}
n_c \sim 1/\gamma  \;\; [ \approx 1/\mu(I)\;\; \text{ for small } \;\; \mu(I)].
\end{equation} 
This scaling
can be obtained from the definition of $n_c$ by assuming a weak dependence of $b/a$  on~$\gamma$. Equation~(\ref{eq.nc}) 
implies that the exponential decay is always dominant 
for small~$I$ and that hyperbolic saddles and
c-measures can be effectively defined for intermediate times $n^*<n<n_c$ (see Fig. \ref{fig4}b). 
An important difference between the manifolds of the hyperbolic and nonhyperbolic
components is that they  attract/repel exponentially and sub-exponentially,
respectively.
The picture that emerges is that 
$\rho_0$ relaxes to the non-hyperbolic component only after, and mainly via, the hyperbolic one. 
Therefore, $b/a$ hardly depends on $\rho_0$ and, consequently, neither does~$n_c$.
This explains scaling~(\ref{eq.nc}) and also why the absolute value of~$n_c$ is approximately the same for escape and recurrence (see the upper inset of Fig. \ref{fig3}). 

In summary, 
we have presented a new interpretation of recurrences based on 
the theory of transient chaos. 
In
particular, we have expressed the exponential relaxation of the recurrence
time distribution in terms of properties of 
chaotic saddles. In addition,
in Hamiltonian systems with mixed phase space we could properly order
the exponential and power-law decays, respectively, to hyperbolic and
non-hyperbolic components of the saddle. Our results hold in continuous time
and in any dimension, being thus valid, e.g., in classical many particle systems.

So far we have emphasized that Poincar\'e recurrences can be viewed as
an escape problem. 
However, in view of the correspondence between the two problems
presented here, also the results on recurrences~\cite{Hadyn,maths,maths2,altmann,BKG}
apply to leaked systems.
Furthermore, the split of the chaotic saddle in Hamiltonian
systems with mixed phase space provides a theoretical description with practical applications in problems like resetting in
hydrodynamics~\cite{pierrehumbert} and emission from optical
cavities~\cite{optics}.
 
We are indebted to L.A. Bunimovich,  G. Gy\"orgyi, H. Kantz, and A. Pikovsky for useful discussions.
This research was supported by the OTKA grant T047233.


\begin{thebibliography}{10}
\bibitem{notenew} Here we assume that map M is invertible. For noninvertible
  maps the invariant set is a chaotic repellor.
\end{thebibliography}
\end{document}